\documentclass{epl2} 

\usepackage{amsmath}
\usepackage{amsfonts}
\usepackage{amssymb}
\usepackage[T1]{fontenc}
\usepackage{subfig}
\usepackage[percent]{overpic}
\usepackage{float}

\title{Immunization strategy for epidemic spreading on multilayer
  networks.}

\author{C. Buono\inst{1} \and L. A. Braunstein\inst{1,2}}
\shortauthor{C. Buono \etal}

\institute{ \inst{1} Instituto de Investigaciones F\'isicas de Mar del
  Plata (IFIMAR)-Departamento de F\'isica FCEyN-UNMDP-CONICET Funes
  3350 (7600) Mar del Plata, Argentina.\\ \inst{2} Center for Polymer
  Studies, Boston University, Boston, Massachusetts 02215, USA}


\abstract{In many real-world complex systems, individuals have many
  kind of interactions among them, suggesting that it is necessary to
  consider a layered structure framework to model systems such as
  social interactions. This structure can be captured by multilayer
  networks and can have major effects on the spreading of process that
  occurs over them, such as epidemics. In this Letter we study a
  targeted immunization strategy for epidemic spreading over a
  multilayer network. We apply the strategy in one of the layers and
  study its effect in all layers of the network disregarding
  degree-degree correlation among layers. We found that the targeted
  strategy is not as efficient as in isolated networks, due to the
  fact that in order to stop the spreading of the disease it is
  necessary to immunize more than the $80 \%$ of the
  individuals. However, the size of the epidemic is drastically
  reduced in the layer where the immunization strategy is applied
  compared to the case with no mitigation strategy. Thus, the
  immunization strategy has a major effect on the layer were it is
  applied, but does not efficiently protect the individuals of other
  layers.}

\begin{document}

\maketitle

\section{Introduction}

The new insights in the complex networks analysis, is no further
considering networks as isolated entities, but characterizing how
networks interact with other networks and how this interaction affects
processes that occurs on top of them. A system composed by many
networks is called \emph{Network of Networks\/} (NoN), a terminology
introduced a few years ago \cite{jia_02,Gao_12,Gao_01,Val13}. In NoN,
there are connectivity links within each individual network, and
external links that connect each network to other networks in the
system. A particular class of Network of Networks in which the nodes
have multiple types of links across different \emph{layers\/}
\cite{Lee_12,Brummitt_12,Gomez_13,Kim_13,Cozzo_12,GoReArFl12,Kiv_13},
are called \emph{Multiplex or Multilayer Networks} \cite{Boc_14}. The
multiplex network approach has proven to be a successful tool in
modeling a number of very wide real-world systems, such as the Indian
air and train transportation networks \cite{Hal_14} and the
International Trade Network \cite{Bar_10,Bar_11}.

In the last couple of years, the study of the effect of multiplexity
of networks in propagation processes such as epidemics has been the
focus of many recent researches
\cite{Dickison_12,Mar_11,Yag_13,Cozzo_13,Zhe_13,Boc_14}. In
Ref~\cite{Buo_14} the research concentrated in the propagation of a
disease in partially overlapped multilayer networks, because the fact
that individuals are not necessarily present in all the layers of a
society impacts the propagation of the epidemic.  For the epidemic
model they used the susceptible-infected-recovered (SIR) model
\cite{Bailey_75,Colizza_06,Colizza_07} that describes the propagation
of non recurrent diseases for which ill individuals either die or,
after recovery, become immune to future infections.  In the SIR model
each individual of the population can be in one of three different
states: Susceptible, Infected, or Recovered. Infected individuals
transmit the disease to its susceptible neighbors with a probability
$\beta$ and recover after a fixed time $t_r$. The spreading process
stops when all the infected individuals are recovered. The dynamic of
the epidemic is controlled by the transmissibility $T$, that is the
effective probability that the disease will be transmitted across any
given contact. As in the SIR model an individual cannot be reinfected,
the disease spreads through branches of infection that have a local
tree-like structure, and thus, this model, can be described using
branching theory approach within a generating function formalism
\cite{Dun_01,New_03} that holds in the thermodynamic limit.  In
\cite{Buo_14}, they found, theoretically and via simulations, that in
the partially overlapped multiplex network, the epidemic threshold
decreases as the overlapped fraction between layers increases, due to
the fact that, when the overlapping between layers increases, the
number of paths the disease can take increases.  They also found that
in the limit of small overlapping fraction, the epidemic threshold is
dominated by the most heterogeneous layer, this effect could have
important implications in the implementation of mitigation strategies.

In a real context, the immunization strategy in social networks is not
made at random.  It is a well known fact that the bigger spreaders in
social networks are those individuals with higher degrees. Some of the
mitigation strategies used in society nowadays are based in this
phenomena, for example, it is mandatory for all hospital staff to get
the vaccine against flu every year, since they are (in average) the
most connected and exposed individuals in the population.  This
suggests that health agencies always try to immunize those individuals
that have, somehow, more chances to get infected and to propagate the
disease.  Motivated by this, in this Letter we study a strategy in
overlapped multiplex networks where the most connected individuals in
one layer are identified and vaccinated, called targeted immunization
strategy. Those immunized overlapped individuals will remain immunized
in all layers of the network.

\section{Model and Results}

\subsection{Immunization strategy}

In our model we use as the substrate for the epidemic spreading a
multiplex network formed by two layers, called $A$ and $B$, of the
same size $N$, and with degree distribution $P_A(k)$ and $P_B(k)$
which are the probability that a random chosen node in layer $A$ and
$B$ respectively has degree $k$. An overlapping fraction $q$ of {\it
  shared individuals} is active in both layers.

For the targeted immunization strategy, we start by immunizing a
fraction $p$ of the highest connected individuals in layer $A$, and as
we assume no degree correlation between layers, the immunization in
layer $B$ will be at random. Immunized individuals can not be infected
by the disease and will remain in the susceptible state during all the
propagation process.

Let $\psi(k)$ be the probability that a node is not immunized given
that it has degree $k$, then $P_A(k)\psi(k)$ is the probability of a
node in layer $A$ to have degree $k$ and not being immunized, and

\begin{equation}
F_0^A(x) = \sum_{k=k_{min}}^{k_{max}} P_A(k) \psi(k) x^k \; ,
\label{F0}
\end{equation}
is the probability generating function for this distribution
\cite{Dun_01} and $k_{min}$ and $k_{max}$ are the minimum and maximum
values of the degrees.  Note that $F_0^A(1)=1-p$, where $1-p$ is the
fraction of the non immunized individuals in layer $A$.

If we follow a randomly chosen link in layer $A$, the node we reach
has degree distribution proportional to $k P_A(k)$, rather than just
$P_A(k)$, because a randomly chosen link is more likely to lead to a
node with higher degree. Hence the equivalent of Eq. (\ref{F0}) for
such a node is \cite{Dun_01},

\begin{eqnarray}
F_1^A(x) = \frac{\sum_k k P_A(k) \psi(k) x^{k-1}}{\sum_k k P_A(k)} =
\frac{F^{ A \prime}_0(x)}{\langle k_A \rangle} \; ,
\label{F1}
\end{eqnarray}
where $\langle k_A \rangle$ is the average node degree in layer $A$,
and $F^{A \prime}_0(x) = dF^{A}_0(x)/dx$.

We need to define the function $\psi(k)$ that will depend on the
immunization strategy used. For the targeted immunization, in layer
$A$ we immunize a fraction $p$ of the higher degree nodes, thus, there
will be a degree cutoff $k_s$ in that layer such that all individuals
with degree higher than $k_s$, and a fraction $w$ of individuals with
degree $k_s$ in layer $A$ are immunized. Therefore, for this strategy,
$\psi(k)$ is,

\begin{equation}
     \label{Esc}
     \psi(k) = \left\{
	       \begin{array}{ll}
		 0      & \mathrm{if\ } k > k_s \; ;\\ 
		 1      & \mathrm{if\ } k < k_s \; ;\\
		 w      & \mathrm{if\ } k = k_s \; .
	       \end{array}
	       \right.
\end{equation}

The total fraction of immunized individuals $p$ can be written as,

\begin{equation}
p = w P_A(k_s) + \sum_{k=k_s+1}^{k_{max}} P_A(k) \; ,
\end{equation}
using the normalization property of the degree distribution
$\sum_{k=0}^{k_{max}} P_A(k) = \sum_{k=0}^{k_s} P_A(k) +
\sum_{k=k_s+1}^{k_{max}} P_A(k) = 1$, we can write $w$ as,

\begin{equation}
w=\frac{p-1+\sum_{k=0}^{k_{max}} P_A(k)}{P_A(k_s)} \; .
\end{equation}

In layer $B$, there is not a direct immunization strategy, however,
the overlapped individuals that were immunized in layer $A$, will be
also immunized in layer $B$ but at random. Thus, there is a fraction
$p q$ of random immunized individuals in layer $B$.

\subsection{Propagation process over the immunized multiplex network}

After the immunization strategy takes place, we start the propagation
process by infecting one randomly chosen susceptible (non immunized)
individual in layer $A$. The spreading process then follows the SIR
dynamics in both layers, and the disease spreads through branches of
infection. We assume that the transmissibility is the same in both
layers and thus all individuals in the system spread equally. The
overlapped nodes in both layers have the same state because they
represent the same individuals.

One parameter that contains all the information about the branching
process is the probability $Q_i$, that choosing a random selected
link, it does not leads to the infinite branch of infected individuals
in layer $i$, with $i=A,B$. The probabilities $Q_A$ and $Q_B$
satisfies the following self consistent equations,

\begin{eqnarray}
&Q_A& = 1 - F^A_1(1) + (1-q) \; F^A_1(1-T+TQ_A) + \nonumber \\ &  & + \; q \; F^A_1(1-T+TQ_A) \;
G^B_0(1-T+TQ_B) \; , \label{QA}\\ 
&Q_B& = pq + (1-q) \; G^B_1(1-T+TQ_B) + \nonumber \\ & & + \; q\; G^B_1(1-T+TQ_B) \; F^A_0(1-T+TQ_A) \; ,
\label{QB}
\end{eqnarray}
were $ G_0^B (x) = \sum_{k=k_{min}}^{k_{max}} P_B (k) x^k $ is the
generating function of the probability to reach a node with degree $k$
in layer $B$ and $G_1^B (x)= \sum_{k=k_{min}}^{k_{max}} \frac{k
  P_B(k)}{\langle k_B \rangle} x^{k-1}$ is the generating function for
the probability to reach a node of degree $k$ in layer $B$ by
following a random chosen link.

Equation (\ref{QA}) has three terms, since the probability $Q_A$ to
not reach the infected branches following a random chosen link in
layer $A$, can be written as the probability that an immunized individual
is reached ($1-F^A_1(1)$), plus the conditional probability that the
reached individual does not have spread the disease given that it is
not immunized. This last conditional probability is split into two
terms, depending if the reached individual is one of the $q$
overlapped fraction or not. If the individual is only present in one
layer with probability $1-q$, the branch will never reach layer $B$
while if the individual is present in both layers with probability
$q$, the branch will reach a node in layer $B$ and can expand through
the $k$ connections of the reached node in that layer. An analogous
interpretation can be made for equation (\ref{QB}).

The solution of the system (\ref{QA}) and (\ref{QB}) above is given by
the intersection of $Q_A$ and $Q_B$. In the criticality, this
intersection can be derived by solving the equation $|J-I|=0$, where
$|\;|$ denotes the determinant, $I$ is the identity and $J$ is the
Jacobian matrix of the system of equations (\ref{QA}) and (\ref{QB}),
which elements are $J_{ij} = \partial Q_{i} / \partial Q_{j}$, with
$i= A,B$ and $j=A,B$. The Jacobian has to be evaluated in $Q_A=Q_B=1$,
since at criticality the disease does not spread and there are no
branches of infection. There are two different eigenvalues for each
one of the possible solutions of the system. The stability of each
solution can be analyzed by the behavior of the eigenvalues, {\it
  i.e.} sink, source or saddle \cite{All_97}. We find that only one of
the possible solutions is stable and therefore, the epidemic threshold
is given by $T_c(q)\equiv T_c$,

\begin{equation}
T_c =\frac{F_1^{ A \prime}(1)+(\kappa_B - 1)(1-pq) - \sqrt{(F_1^{A
      \prime}(1)-(\kappa_B - 1)(1-pq))^2 + 4q^2 F_1^{A}(1)^2 \langle
    k_A\rangle \langle k_B \rangle}}{2 F_1^{ A \prime}(1) (\kappa_B -
  1)(1-pq) - 2 q^2 F_1^{A}(1)^2 \langle k_A\rangle \langle k_B
  \rangle}\; ,
\label{TcT}
\end{equation}
where $F_1^{A \prime}(1)= d F_1^A(x) / d x|_{x=1}$ and
$T_c=1/(\kappa-1)$ where $\kappa$ is the total branching factor of the
multilayer networks.

\vspace{1cm}
\begin{figure}[th]
\centering
 \includegraphics[scale=0.35]{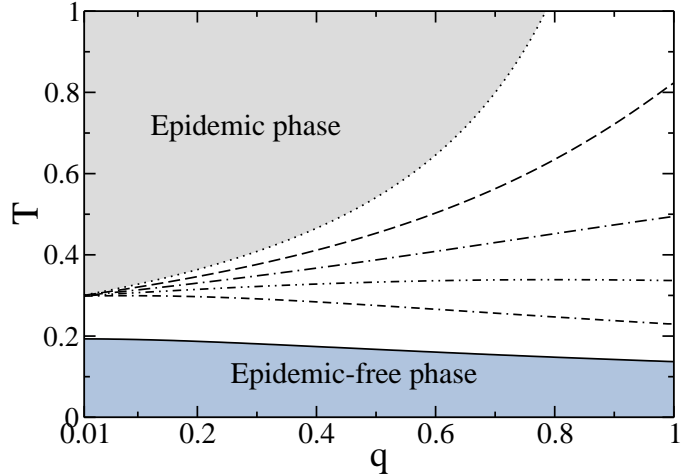}
 \caption{(Color online) Plane $T-q$ for the SIR model in the
   multiplex network, when the targeted immunization strategy is
   applied, for different values of the immunized fraction $p$. Both,
   layer $A$ and $B$, has power law degree distributions $P_{A/B} \sim
   k^{-\gamma_{A/B}}$ with $\gamma_A=2.5$ and $\gamma_B=3.5$ with
   $k_{min}=2$ and $k_{max}=250$. The lines denote the theoretical
   values of $T_c$ for different values of $q$ obtained numerically
   from equation (\ref{QA}) and (\ref{QB}). From top to bottom
   $p=0.9;0.7;0.5;0.3;0.1;0.01$. Above the lines the system is in the
   epidemic phase for each value of $p$, and below it is in the
   epidemic-free phase where the disease dies out.}
\label{diag}
\end{figure}

In Fig. \ref{diag} we plot the plane T-q obtained from
Eq. (\ref{TcT}), for different values of $p$. We use a power law
degree distribution $P_{A/B} \sim k^{-\gamma_{A/B}}$ in both layers
with exponents $\gamma_A=2.5$ and $\gamma_B=3.5$ in layer $A$ and $B$
respectively, where $k_{min}=2$ and $k_{max}=250$ are the minimum and
maximum connectivity. Note that layer $A$ in which the immunization is
applied is the most heterogeneous layer, however similar results are
found using different degree distributions on each layer. The lines
represent $T_c$ for many values of $p$, above the lines there is an
epidemic phase and below $T_c$ only outbreaks exists (non-epidemic
phase). Fig. \ref{diag} shows that $T_c$ has different behaviors with
$q$ depending on the value of $p$.

For $q=0$ (not shown) the critical threshold corresponds to an
isolated layer in which the disease starts, {\it i.e.} layer $A$ and
where the critical threshold is given by $T_c = 1/F_1^{ A \prime}(1)$,
where $F_1^{ A \prime}$ is not the branching factor of layer $A$, but
gives a measure of the heterogeneity of the layer. For $q \to 0$ the
process is dominated by the most heterogeneous layer \cite{Buo_14},
therefore, the epidemic threshold converges to the threshold of that
layer. In Fig. \ref{diag} we can see that for $p=0.01$ and $q\to 0$,
$T_c = 1/F_1^{ A \prime}(1)$, due to the fact that the most
heterogeneous layer is $A$, while for $p \geq 0.1$, $T_c = 1/ G_1^{B
  \prime}(1) = 1/\kappa_B - 1$ were $\kappa_B$ is the branching factor
of layer $B$.

From the phase diagram (see Fig~\ref{diag}) we can see that for $p <
0.2$, $T_c$ decreases with $q$, this behavior agrees with the expected
non-immunized behavior, since as $q$ increases, the total branching
factor of the network increases and thus $T_c$ decreases
\cite{Buo_14}. For $p > 0.2$, $T_c$ increases with $q$, due to the
fact that layer $A$ gets fragmented and the disease spreads through
layer $B$, and as $q$ increases, the fraction $p \;q$ of the random
immunized individuals in layer $B$ increases hindering the spreading
through layer $B$ and thus, $T_c$ increases with $q$. When the
fraction of immunized individuals $p \;q > 0.72$, layer $B$ also gets
fragmented, and the disease can not spread at all, thus, the epidemic
regime disappears as shown in figure \ref{diag} for $p=0.9$ and
$q\gtrsim 0.79$.  We can understand this behavior using percolation
theory, for the targeted percolation process. For this process it was
found that \cite{Coh_10} the critical value of the percolation
fraction $\hat{p_c}$ in scale free networks with exponent $\gamma =
2.5$ is $\hat{p_c} \approx 0.2$ such that for $p > \hat{p_c}$ the
networks is fragmented and for $p < \hat{p_c}$ there is a giant
connected cluster, that is also the critical threshold corresponding
to the targeted immunization strategy in layer $A$. In layer $B$,
there is a random immunization equivalent to a random percolation
process, for which the critical value of percolation fraction
$\hat{p_c}$ in scale free networks with exponent $\gamma=3.5$ is
$\hat{p_c} \approx 0.72$ \cite{Coh_10} and corresponds to the critical
threshold due to the random immunization strategy in layer $B$.
Despite that the targeted immunization strategy is the best strategy
to stop propagation in isolated networks, in overlapped multiplex
networks it is not as efficient due to the fact that the threshold is
dominated by the most heterogeneous network. From the phase diagram we
can observe that in order to suppress the epidemic phase one has to
immunize more that the $80 \%$ of the population in layer $A$. Thus
even if network $A$ is fragmented ($p > 0.2$) the disease can still
propagating in network $B$ which is more heterogeneous than the
fragmented layer $A$.  Notice that layer $B$ is not fragmented for
$p\;q < 0.72$.

However, even if it is hard to stop the epidemic, its size can be
drastically reduced compared to the case where no strategy is applied.
The size of the epidemic can be computed as the total number of
recovered individuals in the final state of the epidemic, and is given
by,

\begin{eqnarray}
R_A = & q & \left [ 1 - p - F_0^A(1-T+T\;Q^*_A) G_0^B(1-T+T\;Q^*_B) \right ] + \nonumber \\ 
& + & (1-q) \left [ 1 - p - F_0^A(1-T+T\;Q^*_A) \right ] \label{RA}  \; ;\\ 
R_B = & q & \left [ 1 - p - F_0^A(1-T+T\;Q^*_A) G_0^B(1-T+T\;Q^*_B) \right ] + \nonumber \\
& + & (1-q) \left [ 1 - G_0^B(1-T+T\;Q^*_B) \right ] \; , \label{RB}
\end{eqnarray}
where $Q^*_A$ and $Q^*_B$ are the non trivial solutions of
Eqs. (\ref{QA}) and (\ref{QB}) for $T \gtrsim T_c$.
 
\begin{figure}[th]
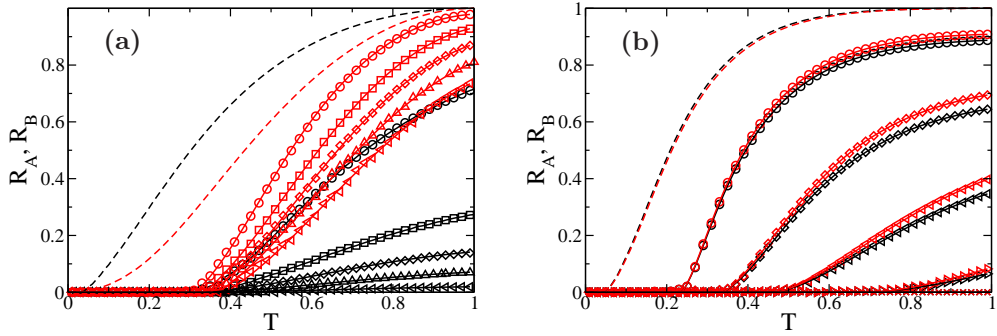

\centering
 \begin{overpic}[scale=0.25]{fig2a.eps}
    \put(20,60){{\bf{(a)}}}
  \end{overpic}\hspace{0.5cm}
  \begin{overpic}[scale=0.25]{fig2b.eps}
    \put(20,60){\bf{(b)}}
  \end{overpic}\vspace{0.5cm}
 \caption{(Color online) Fraction of recovered individuals in the
   final state of the epidemics for layer $A$, $R_A$ (black), and
   layer $B$, $R_B$ (red), as a function of $T$. Both layers have
   power law degree distributions $P_{A/B} \sim k^{-\gamma_{A/B}}$
   with $\gamma_A=2.5$ and $\gamma_B=3.5$ for layer $A$ and $B$
   respectively. Lines denote the theoretical results obtained from
   Eqs. (\ref{RA}) and (\ref{RB}) while symbols denote numerical
   simulation results for layer size $N=10^5$ and over $10^5$ network
   realization.  a) $q=0.2$ the dashed lines correspond to the case
   without immunization strategy $p=0$, and $p=0.1;0.3;0.5;0.7;0.9$
   from top to bottom b) $q=0.9$, the dashed lines correspond to the
   case without immunization strategy $p=0$, and the full lines and
   symbols corresponds to $p=0.1;0.3;0.5;0.7;0.9$ from top to bottom.}
\label{r01}
\end{figure}

In figures \ref{r01} (a) and (b) we plot the results of $R_A$ and
$R_B$ as a function of $T$, obtained both theoretically from
Eqs. (\ref{RA}) and (\ref{RB}) and from the numerical simulation. We
found a good agreement between the theoretical results (lines) and the
numerical simulations (symbols). In Fig. \ref{r01} (a) we show the
results for $q=0.2$ and different values of $p$. If we compare the
results with and without strategy (dashed line) we can see that the
immunization strategy not only affects the epidemic threshold, but
also decreases the impact of the disease in both layers, since at
fixed $T$, both $R_A$ and $R_B$ decreases with $p$. For $p \geq 0.2$
layer $A$ gets fragmented and the disease never reaches more than $30
\%$ of the individuals in that layer, however the impact of the
disease in layer $B$ is significant and even for $p = 0.9$, more that
$60 \%$ of the individuals in layer $B$ can be infected. Therefore,
for an overlapping fraction between layers $q=0.2$, the immunization
strategy has a major effect on the layer were it is applied, but does
not protect the individuals of other layers.

In Fig. \ref{r01} (b), we show $R_A$ and $R_B$ for $q=0.9$, and we can
see that even though one needs to immunize more than $80 \%$ of the
individuals to suppress the epidemic phase, the impact of the disease
in both layers decreases significantly with $p$. We compare $R_A$ and
$R_B$ with and without strategy (dashed line) and see that, in this
regime, due to the high overlapping between layers, the effect of the
immunization strategy on layer $B$ is stronger, difficulting the
propagation through that layer. Therefore when the overlapping between
layers is high, the strategy is more efficient to protect the
individuals of the whole network.

\vspace{1cm}

\section{Conclusions}  

In this Letter we study, theoretically and via simulations, a targeted
immunization strategy for epidemic spreading in a partially overlapped
multiplex network composed by two layer with an overlapping fraction
$q$. We immunize a fraction $p$ of individuals in one layer of the
network and study how this process affects the propagation of the
disease through all layers. We found that the branching theory gives a
good approach of the phenomena. For $q \to 0$ the critical threshold
of the epidemic is dominated by the threshold of the most
heterogeneous layer for all $p$. When $p$ is smaller than the critical
percolation threshold of layer $A$, $T_c$ decreases with $q$, as in
the non immunized model presented in \cite{Buo_14}. When $p$ is above
the criticality of layer $A$, this layer gets fragmented and thus the
epidemic can only spread through layer $B$. The fraction of immunized
individuals in layer $B$ is $pq$, thus as $q$ increases $T_c$
increases. This behavior holds until the fraction $p q$ exceeds the
critical percolation threshold of layer $B$. Above this threshold
layer $B$ also gets fragmented and thus, the disease can not spread at
all, suppressing the epidemic phase. This regime can only be reached
if one immunizes more than $80 \%$ of individuals. However, even if it
is hard to stop the epidemic, its size can be drastically reduced
compared to the case where no strategy is applied. We found that the
immunization strategy has a major effect on the layer were it is
applied, but does not efficiently protect the individuals of other
layers.

Real networks of networks such as the world wide port network and a
world wide airport network \cite{Par_10} have assortative
degree-degree correlation between networks, {\it i.e.} biggest
airports are connected with bigger ports. In order to have a realistic
scenario we should consider degree-degree correlation between
layers. If the correlation is assortative, then the immunization
strategy in layer $B$ will be also targeted and thus the strategy
would be more effective since layer $B$ will get fragmented easier. In
a future work, we will study deeply the effects of correlations
between layers in the epidemic spreading and immunization strategies
in multilayer networks.

After this letter was submitted a similar strategy was
  published by D. Zhao {\it et. al} in \cite{Zha_14}.

\acknowledgments This work was financially supported by UNMdP and
FONCyT (Pict 0429/2013). The authors thank Lucas D. Valdez for his
useful comments and discussions.

\bibliographystyle{eplbib.bst}
\bibliography{Buono}

\begin{thebibliography}{10}
\expandafter\ifx\csname url\endcsname\relax\def\url#1{\texttt{#1}}\fi

\bibitem{jia_02}
\Name{Gao J., Buldyrev S.~V., Havlin S. \and Stanley H.~E.} \REVIEW{Phys. Rev.
  Lett.}{107}{2011}{195701}.

\bibitem{Gao_12}
\Name{Gao J., Buldyrev S.~V., Stanley H.~E. \and Havlin S.} \REVIEW{Nature
  Physics}{8}{2012}{}.

\bibitem{Gao_01}
\Name{Dong G., Gao J., Du R., Tian L., Stanley H.~E. \and Havlin S.}
  \REVIEW{Phys. Rev. E}{87}{2013}{052804}.

\bibitem{Val13}
\Name{Valdez L.~D., Macri P.~A., Stanley H.~E. \and Braunstein L.~A.}
  \REVIEW{Phys. Rev. E}{88}{2013}{050803(R)}.

\bibitem{Lee_12}
\Name{Lee K.-M., Kim J.~Y., Cho W.~K., Goh K.-I. \and Kim I.-M.} \REVIEW{New J.
  Phys.}{14}{2012}{033027}.

\bibitem{Brummitt_12}
\Name{Brummitt C.~D., Lee K.-M. \and Goh K.-I.} \REVIEW{Phys. Rev.
  E}{85}{2012}{045102(R)}.

\bibitem{Gomez_13}
\Name{G{\'o}mez S., D{\'i}az-Guilera A., G{\'o}mez-Garde{\~n}es J.,
  P{\'e}rez-Vicente C.~J., Moreno Y. \and Arenas A.} \REVIEW{Phys. Rev.
  Lett.}{110}{2013}{028701}.

\bibitem{Kim_13}
\Name{Kim J.~Y. \and Goh K.-I.} \REVIEW{Phys. Rev. Lett.}{111}{2013}{058702}.

\bibitem{Cozzo_12}
\Name{Cozzo E., Arenas A. \and Moreno Y.} \REVIEW{Phys. Rev.
  E}{86}{2012}{036115}.

\bibitem{GoReArFl12}
\Name{G{\'o}mez-Garde{\~n}es J., Reinares I., Arenas A. \and Floria L.~M.}
  \REVIEW{Nature Scientific Reports}{10.1038}{2012}{srep00620}.

\bibitem{Kiv_13}
\Name{Kivel{\"a} M., Arenas A., Barthelemy M., Gleeson J.~P., Moreno Y. \and
  Porter M.~A.} \Book{{Multilayer Networks}} http://arxiv.org/abs/1309.7233
  (2013).

\bibitem{Boc_14}
\Name{Boccaletti S., Bianconi G., Criado R., del Genio C.,
  G{\'o}mez-Garde{\~n}es J., Romance M., Sendi{\~n}a-Nadal I., Wang Z. \and
  Zanin M.} \REVIEW{Physics Reports}{544}{2014}{1}.

\bibitem{Hal_14}
\Name{Halu A., Mukherjee S. \and Bianconi G.} \REVIEW{Phys. Rev.
  E}{89}{2014}{012806}.

\bibitem{Bar_10}
\Name{Barigozzi M., Fagiolo G. \and Garlaschelli D.} \REVIEW{Phys. Rev.
  E}{81}{2010}{046104}.

\bibitem{Bar_11}
\Name{Barigozzi M., Fagiolo G. \and Mangioni G.} \REVIEW{Physica
  A}{390}{2011}{2051}.

\bibitem{Dickison_12}
\Name{Dickison M., Havlin S. \and Stanley H.~E.} \REVIEW{Phys. Rev.
  E}{85}{2012}{066109}.

\bibitem{Mar_11}
\Name{Marceau V., No{\"e}l P., H{\'e}bert-Dufresne L., Allard A. \and Dub{\'e}
  L.~J.} \REVIEW{Phys. Rev. E}{84}{2011}{026105}.

\bibitem{Yag_13}
\Name{Yagan O., Qian D., Zhang J. \and Cochran D.} \REVIEW{IEEE JSAC Special
  Issue on Network Science}{31}{2013}{1038}.

\bibitem{Cozzo_13}
\Name{Cozzo E., Ba{\~n}os R.~A., Meloni S. \and Moreno Y.} \REVIEW{Phys. Rev.
  E}{88}{2013}{050801(R)}.

\bibitem{Zhe_13}
\Name{{Wang Zhen}, {Szolnoki Attila} \and {Perc Matjaz}} \REVIEW{Sci.
  Rep.}{3}{2013}{2470}.

\bibitem{Buo_14}
\Name{Buono C., Zuzek L. G.~A., Macri P.~A. \and Braunstein L.~A.} \REVIEW{PLoS
  ONE}{9}{2014}{e9220}.

\bibitem{Bailey_75}
\Name{Bailey N. T.~J.} \Book{{The Mathematical Theory of Infectious Diseases}}
  (Griffin, London) 1975.

\bibitem{Colizza_06}
\Name{Colizza V., Barrat A., Barthlemy M. \and Vespignani A.} \REVIEW{Proc.
  Natl. Acad. Sci. USA}{103}{2006}{2015}.

\bibitem{Colizza_07}
\Name{Colizza V. \and Vespignani A.} \REVIEW{Phys.Rev.Lett.}{99}{2007}{148701}.

\bibitem{Dun_01}
\Name{Callaway D., Newman M. E.~J., Strogatz S.~H. \and Watts D.~J.}
  \REVIEW{Phys. Rev. Lett.}{85}{2000}{5468}.

\bibitem{New_03}
\Name{Newman M. E.~J., Strogatz S.~H. \and Watts D.~J.} \REVIEW{Phys. Rev.
  E}{64}{2001}{026118}.

\bibitem{All_97}
\Name{Alligood K.~T., Sauer T.~D. \and Yorke J.~A.} \Book{{CHAOS: An
  Introduction to Dynamical Systems}} (Springer) 1997.

\bibitem{Coh_10}
\Name{Cohen R. \and Havlin S.} \Book{{Complex Networks: Structure, Robustness
  and Function}} (Cambridge University Press) 2010.

\bibitem{Par_10}
\Name{Parshani R., Rozenblat C., Ietri D., Ducruet C. \and Havlin S.}
  \REVIEW{EPL}{92}{2010}{68002}.

\bibitem{Zha_14}
\Name{Zhao D., Wang L., Li S., Wang Z., Wang L. \and Gao B.} \REVIEW{PLOS
  ONE}{9}{2014}{e112018}.

\end{thebibliography}

\end{document}